# SEMANTIC WEB APPROACH TOWARDS INTEROPERABILITY AND PRIVACY ISSUES IN SOCIAL NETWORKS

Dr V.Kiran Kumar

Department of Computer Science Dravidian University, Kuppam

*ABSTRACT*

*The Social Web is a set of social relations that link people through World Wide Web. This Social Web encompasses how the websites and software are designed and developed to support social relations. The new paradigms, tools and web services introduced by Social Web are widely accepted by internet users. The main drawbacks of these tools are it acts as independent data silos; hence interoperability among applications is a complex issue. This paper focuses on this issue and how best we can use semantic web technologies to achieve interoperability among applications.*

*KEYWORDS*

*Social Web, Semantic Web, interoperability, Web*

## I. INTRODUCTION

The main aim of Social Web is make relations among people on web. The Social Web encompasses how the websites and software are designed and developed to support social relations. Web 2.0 has introduced such applications to make relations among people. The examples are Facebook, twitter, Myspace, YouTube etc., One of the drawback of these social websites is that these tools acts as independent data silos. This paper has explains the issues relating to the social web and discusses about how best we can solve this with the help of semantic web technologies.

The rest of the paper is organized as follows. Section 2 gives an introduction to Social Web and the technologies used. Section 3 discusses about the privacy and interoperability issues relating to Social Web. Section 4 discusses about an introduction to semantic web. Section 5 deals with how best we can use Semantic Web technologies to achieve interoperability in Social Web. Section 6 ends with a brief conclusion of this paper.

## II. LITERATURE REVIEW

Li Ding, Tim Finin and Anupam Joshi in their paper on "Analyzing social networks on the Semantic Web", had described a research on integrating social ontologies and extracting social networks on the semantic web. They had focused on how far various ontologies are being used in social networks with empirical analysis. The Swoogle ontology directory shows that the *foaf:person* currently has one million instances spread over by more than 45000 documents.

Uldis Bojaris, Alexandra Passant, John G.Breslin and Stefen Decker had published a paper on "Social Network and Data Portability using semantic web technologies". They discussed various challenges regarding social networks and data portability and how the semantic web can be used for achieving the goal. They used Friend of Friend (FOAF) ontology to represent identity and distributed social networks and Semantically-Interlinked online communities (SIOC) ontology can be used to represent data on social networks of different users from different sites.

Till Plumbaum, Songxuan Wu, Ernesto William De Luca and Sahin Albayarak had studied 17 social network applications to share user data across different applications in their article on "User Modelling for the Semantic Web". They proposed a Social Web User Model (SWUM) that is fitted to the needs of Social Web to cater interoperability issue. After thorough analysis on social networks, they pointed certain dimensions like personal characteristics, interests, knowledge behavior and so on.

John G.Breslin, Alexandre Passant and Denny Vrandec had briefly discussed about social web and Semantic Web in their article "Social Semantic Web". They explained in detail explanations regarding the issues in social web, various ontologies available for semantic and how to bridge the gap between social web and semantic web technologies. They explained various semantic web applications like semantic blog, semantic wiki, semantic microblogging etc., Further they discussed about trust and privacy issues with semantic web.

## III. SOCIAL WEB

The Social Web is a set of social relations that link people through the World Wide Web[1]. The term Social Web was coined by Howard Rheingold in 1996 in his article on "Electric Minds"[2]. The most popular and mostly used interactions between people are through telephones. The invention of Web 2.0 applications had changed the social relations among people in a tremendous manner. The second generation applications had developed number of software applications which can provide social relations among people. The examples web 2.0 applications are Wikipedia, blogs, folksonomies, social networking sites, video sharing sites and mashups etc., The main characteristics of Second Generation Websites include community, mashups and AJAX (Asynchronous JavaScript with XML).

Community Websites refers to a group of individuals who share a common interest via e-mail, blogs and chat rooms and so on. Mashups are services from different sites that can be pulled together in order to experience the data in a novel and enhanced way. AJAX is a group of interrelated web development techniques used on client side to send or retrieve data from server asynchronously.

The social networking sites (SNS) and content sharing sites are most popular in a way of providing social relations among people. The Social Networking Sites (SNS) include Face book, Friend setter, orkut, LinkedIn and MySpace. The social networking site is used to describe any Web Site that enables users to create public profiles within the Web Site and form relationships with other users of the same Web Site. The most popular Social Networking Site (SNS) is Face book, where a user can create his own profile with all details like date of birth, place of birth, qualification, working particulars etc., Based on the details given by the user the Web Site searches for the people for correct match like place of birth, course studies on the same year, employees details and so on. This was founded by Mark Zuckerberg in February 2004. Similarly LinkedIn is also most popular and provides professional relations among people.

Content sharing sites are another mode of sharing content among people. The most popular content sharing sites include YouTube (Video Sharing), Flickr (Images sharing) and Last.fm

(Music files sharing) so on. The history of YouTube began on February 14, 2005 when three former PayPal employees activated the Internet domain name "YouTube.com" and started to create a video-sharing website on which users could upload, share, and view videos[3]. This was the most popular video sharing website where it includes cinema teasers, educational videos for kids as well as adults, adult videos and so on.

## IV. ISSUES WITH THE SOCIAL WEB

- **Privacy**

Today it's frequent to hear people saying "everybody is on Facebook". The Facebook Company is also designing applications to run even on mobile devices. Hence, everybody is using Facebook to share information with others. But many people are giving their information on net without knowing how to make it as private. Hence, everybody can see their personal details if it is in public. One of the major problems in Social Networking Sites are "Cyberstalking". This is nothing but the repeated use of electronic communications to harass or frighten someone by sending mails, uploading videos etc[4]. Some applications are developed to get the person's information on the web. Creepy is a Geo location tool which can track person's location using photos uploaded on Face book or Twitter. With this application one can tract person's location when he uploaded images on the web.

- **Interoperability**

The major problems of Social Networking Sites (SNS) are isolation. When a person is registered with one Website cannot access other Social Networking Website with same registration. For example, when a person registered with Facebook cannot access LinkedIn because each website is designed for their own promotion and commercial aspects. If a person wishes to get feedback of a product, then one website contains about the image and other websites may contains information about reviews, configurations and so on. For example, you want to purchase a smart phone, and then you may find comments about the product in Twitter, YouTube and so on. Hence, you need to login to all famous Web sites for getting of information about the product. This is due to lack of interoperability among different social networking sites.

Another important aspect of Social Networking Sites (SNS) is binding users to their website. The person cannot move from one social networking site to another without losing the previously added and maintained information. The social networking sites are locking the users by not providing interoperability. Recently, some websites are accepting Face book id to login in to their website, hence this problem can be minimized at some extent.

The technical reason for not providing interoperability by different social networking websites is due to lack of common standards for knowledge and information sharing. Social Networking Sites (SNS) like blogs, Wikipedia and forums contains much information. The reusing of information in Wikipedia with other application on the web is also a big challenge.

## V. SEMANTIC WEB

According to the World Wide Web Consortium (W3C), "The Semantic Web provides a common framework that allows data to be shared and reused across application, enterprise, and community boundaries."[5] The term was coined by Tim Berners-Lee for a web of data that can be processed by machines[6]. Fig.1 represents Semantic Layer with different technologies are used in designing Semantic Web applications. The primary purpose of these languages is to represent machine-understandable information and to support interoperability between applications on web. Once we add semantics to the website, we can design semantic web applications for the users to use. Uniform Resource Identifier (URI) represents any resource on the web with unique name. The key technologies include Resource Description Framework (RDF), Resource Description Framework Schema (RDFS) and Ontology Web Language (OWL).

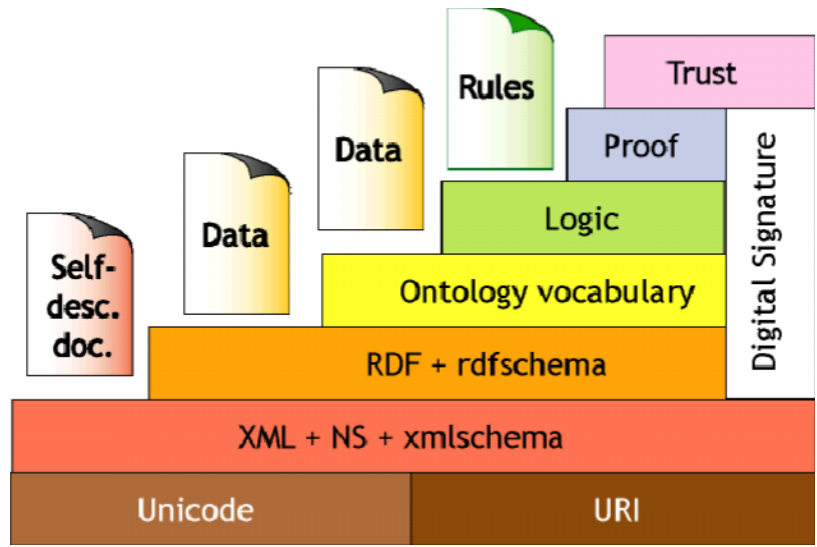

Fig.1 Semantic Layer Cake

## VI. SEMANTIC WEB AND SOCIAL WEB

Semantic Web aims to provide flexible standards for information exchange and interoperability. To provide common standards for different applications it uses vocabularies. Vocabularies define the concepts and relationships (also referred to as "terms") used to describe and represent an area of concern[7]. The main role of vocabularies used in semantic web is to clear ambiguity among different applications. For example, a bookseller wants to integrate data from different publishers. Then different publishers may use different terms for data storage i.e., one publisher may used as "author" and another publisher may used as "creator". As we know that both terms are equal, to make such equal relation we must use vocabulary. Vocabulary makes equal relation between author and creator. Vocabulary can also be called as ontology.

Semantic web contains different vocabularies for different purposes. Such are RSS 1.0 (RDF Site Summary), FOAF (Friend of a Friend) and SIOC (Semantically Interlinked Online Communities). With these vocabularies, social data can be represented using shared and common models and therefore it become more interoperable and portable across applications.

RDF Site Summary (RSS) is a lightweight multipurpose extensible metadata and description format. Metadata is nothing but data about data. In HTML, we use to give information about a webpage in <head> section. This is simple example of metadata, which is meant for machine process able data. Most of the search engines use these Meta data for ranking the web pages and so on. FOAF (Friend of a Friend) is a machine readable ontology describing persons, their objectives and their relations to other people[8]. Most of the people are describing themselves on the net with the help of FOAF vocabulary and help machines to understand their webpage.

Semantically-Interlinked Online Communities Project (SIOC pronounced as "shock") is a semantic web ontology that provides methods for interconnection discussion methods such as blogs, forums and mailing lists to each other[9]. It is composed of core ontology and a set of modules focusing of integration with other applications. Similarly there are number of vocabularies were described by Semantic Web group, such are Dublin core, TrackBack, Vcard RDF, GoodRelations, DOAP, Music Ontology and so on. With the help of these vocabularies we can integrate applications and provides interoperability among applications on social web.

## VII. CONCLUSION

This paper briefly explains about the Social Web and their technologies used. Web 2.0 has developed many applications concerns with social relations. This paper focuses on the advantages of these applications and at same time discusses about privacy and interoperability issues. The Semantic Web has developed number of vocabularies to achieve interoperability among applications over web. This paper discusses about some of the vocabularies in brief and how best we can use these vocabularies to achieve interoperability among applications. We hope that this paper is useful for researchers who are working on semantic web.